\documentclass[a4paper, 10pt,titlepage]{article}
\makeatletter                                               
\renewcommand{\@cite}[1]{#1}                               
\makeatother                                               
\newcommand{\onlinecite}[1]{\cite{#1}}                     
\newcommand{\supcite}[1]{{\tiny $^\textrm{\cite{#1}}$  }}

\newcommand{\rcite}[1]{Ref.~\onlinecite{#1}}               

\usepackage{amsmath}
\usepackage{epsfig}
\usepackage{psfrag}
\usepackage[latin1]{inputenc}

\newcommand{\fig}[1]{Fig.~\ref{#1}}
\newcommand{\eq}[1]{Eq.~(\ref{#1})} 
\newcommand{\rapp}[1]{Appendix \ref{#1}}
\newcommand{\rsec}[1]{Sec.~\ref{#1}}
\newcommand{\fat}[1]{\mbox{\boldmath$#1$}}
\newcommand{\script}[1]{{\textrm{\scriptsize #1}}}
\newcommand{\su}[1]{_\script{#1}}

\newcommand{\nhat}{\hat{\bf n}}

\begin{document}
\twocolumn[\large {}]            

\noindent
{\bf \Large An intuitive approach to inertial \\ forces and the centrifugal force \\ 
paradox in general relativity}\\[3mm] 
{\bf Rickard M. Jonsson}\\[2mm] 
{\it Department of Theoretical Physics, Physics and Engineering
Physics, Chalmers University of Technology, and G\"oteborg University, 412
96 Gothenburg, Sweden\\[2mm] 
}
E-mail: rico@fy.chalmers.se\\[2mm]
Submitted: 2004-12-09, Published: 2006-10-01\\
Journal Reference: Am. Journ. Phys. {\bf 74} 905\\
\\
{\bf Abstract}. As the velocity of a rocket in a circular orbit near a black hole increases,
the outwardly directed rocket thrust must increase to keep the rocket in its
orbit. This feature might appear paradoxical from a Newtonian viewpoint, but
we show that it follows naturally from the equivalence principle together
with special relativity and a few general features of black holes. 
We also derive a general relativistic formalism of inertial forces for
reference frames with acceleration and rotation. The resulting equation
relates the real experienced forces to the time derivative of the speed and
the spatial curvature of the particle trajectory relative to the reference
frame. We show that an observer who follows the path taken by a free
(geodesic) photon will experience a force perpendicular to the direction of
motion that is independent of the observers velocity. 
We apply our approach to resolve the submarine paradox, which regards
whether a submerged submarine  in a balanced state of rest 
will sink or float when given a horizontal
velocity if we take relativistic effects into account.
We extend earlier treatments of this topic to include spherical oceans and show that 
for the case of the Earth the submarine floats upward if we take the curvature of the ocean into account.

\section{Introduction}
\begin{figure}[h]
\begin{center}
\epsfig{figure=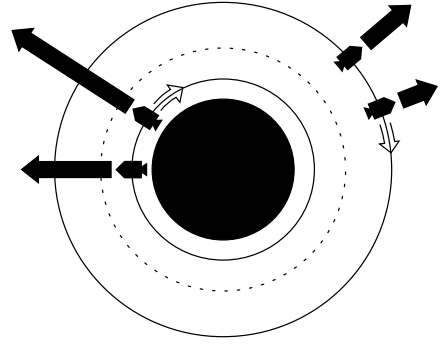,width=6cm}
\caption{
Rockets orbiting a static black hole. The solid arrows correspond to
the force (the rocket thrust) necessary to keep the rocket in circular
orbit. Inside of the photon radius (the
dashed circle), the required force 
increases as the orbital velocity increases. 
}
\label{fig1}
\end{center} 
\end{figure}

Consider a rocket in a circular orbit outside the event horizon of a
black hole.\supcite{hori}
If the orbit lies within the photon radius, the radius
where free photons can move on circular orbits,\supcite{photradius} 
a greater outward rocket thrust is required to keep the rocket in orbit the
faster the rocket moves. However, outside of the photon radius the 
outward thrust decreases as the orbital speed increases just as it would for
a similar scenario in Newtonian mechanics (the thrust will be inward directed
for sufficiently high speeds, see \fig{fig1}). 

Analogous to the situation in Newtonian mechanics we can introduce in general
relativity a gravitational force that is velocity independent. This force is
fictitious (unlike in Newtonian mechanics). We can also introduce a velocity
dependent (fictitious) centrifugal force that together with the
gravitational force balances the real force from the jet engine of the
rocket. By this definition, the centrifugal force is directed inward inside of the photon
radius and directed outward outside of the photon radius. 
This reversal of
the direction of the fictitious centrifugal force is described by the
formalism of optical geometry (see
\rapp{opt}) in which the phenomena has been
discussed.\supcite{marek,orig,carter,mnras1,mnras3,nature} 

Our purpose is not to explain the velocity dependence of the rocket thrust
by analogy with Newtonian theory, and we will use neither gravitational nor
centrifugal forces. Instead we will use the basic principles of relativity to
explain how the real force required to keep an object moving along a
specified path depends on the velocity of the object.

We start by illustrating how the fact that the rocket thrust increases with
increasing orbital speed (sufficiently close to the black hole) follows
naturally from the equivalence principle (reviewed in
\rapp{extraapp}). We do so by first considering an idealized special
relativistic scenario of a train moving relative to an (upward) accelerating
platform. 

We then consider a more general but still effectively
two-dimensional discussion of forces perpendicular to the direction of
motion for motion relative to an accelerated reference frame in special
relativity. In a static spacetime, the reference frame connected to the
static observers behaves locally like an accelerating reference frame in
special relativity and the formalism can therefore be applied also to this case.

We then illustrate how to apply the formalism of this paper to the submarine
paradox.\supcite{submarine} We ask whether a submarine with a density such that
it is vertically balanced when at rest, will sink or float if given a
horizontal velocity and relativistic effects are taken into account.

Next we generalize the formalism of forces and curvature of spatial
paths to include three-dimensional cases, forces parallel to the
direction of motion, and rotating reference frames. The acceleration and rotation of the reference
frame will be shown (as in Newtonian mechanics) to introduce terms that can
be interpreted as inertial (fictitious) forces. By using the equivalence
principle, the formalism can be applied to arbitrary rigid reference frames
in general relativity. We verify the results by comparing with
\rcite{rickinert}.

Although this paper is primarily aimed at readers with a background in
general relativity, the main part assumes only an elementary knowledge of
special relativity together with a knowledge of a few basic concepts of
general relativity. Some of the more important concepts such as curvature of
a spatial path, spatial geometry, geodesics, and the equivalence principle
are reviewed in \rapp{extraapp}. Sections~\ref{3d}--\ref{rotating} are more
specialized.

\section{The train and the platform}\label{train}
\begin{figure}[h]
\begin{center}
\psfrag{v}{$v$}
\psfrag{(a)}{(a)}
\psfrag{(b)}{(b)}
\epsfig{figure=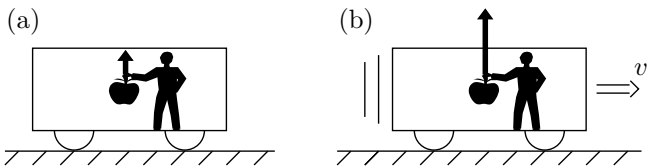,width=8.5cm}
\caption{A train on a platform with a
constant proper acceleration $a$ upward. (a) The train is
at rest; (b) the train is moving relative to the
platform. The force required of a man on the train to keep an
apple at a fixed height is higher when the train moves than when
it is at rest relative to the platform.
} 
\label{fig2}
\end{center} 
\end{figure}

We consider the special relativistic description of a train moving relative
to a platform with proper upward acceleration
$a$.\supcite{proper} The force required by a man on the train to hold an apple
at a fixed height increases as the train speed increases (assuming nonzero
acceleration of the platform) as illustrated in \fig{fig2}. 

To understand this effect we consider the accelerating train as observed
from two inertial systems. The first system $S$ is a system in which
the platform is momentarily ($t=0$) at rest. The second system $S'$
is comoving with the train at the same moment.
The two systems are related to each other by a Lorentz transformation of
velocity $v$, where $v$ is the velocity of the train relative to the
platform along the
$x$-axis.

Relative to $S$ the apple moves to the right and
accelerates upward with acceleration $a$. 
Consider now two physical events at the apple, one at $t=0$ and one at
$t=\delta t$, as observed
from $S$. 
The time separation as observed in $S'$ is (to lowest nonzero order in
$\delta t$) given by $\delta t'=\delta t/\gamma$, where
$\gamma=1/\sqrt{1-v^2/c^2}$ and $c$ is the velocity of light. 
In the following we will use $c$ as the unit of velocity 
so that $v=1$ for photons.\supcite{cett}
The height $\delta h$ separating the two events as observed in $S$ 
equals the corresponding separation $\delta h'$ relative to $S'$. If we denote
the upward acceleration relative to $S'$ by $a'$, we have to lowest
nonzero order in $\delta t$
\begin{subequations}
\begin{align}
\delta h&=a \delta t^2/2 \\
\delta h'&=a'\delta {t'}^2/2\\
\delta h'&=\delta h.
\end{align}
\end{subequations}
From these equations follows that 
\begin{equation}
a' =a \Big( \frac{\delta t}{\delta t'} \Big)^2=a \gamma^2.
\end{equation}
Thus the proper acceleration, that is, the acceleration as observed
from an inertial system momentarily comoving with the apple, 
is greater than the acceleration of the platform
by a factor of $\gamma^2$. The force required to keep the apple of rest mass
$m$ at a fixed height relative to the train is thus given by $F=m
\gamma^2 a$. 

\begin{figure}[h]
\begin{center}
\psfrag{a}{$a$}
\epsfig{figure=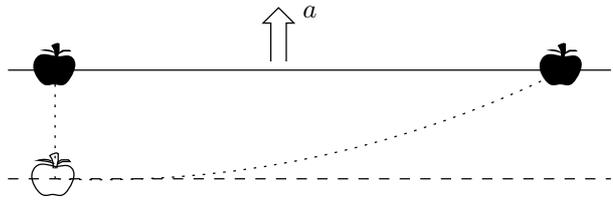,width=8cm}
\caption{Two apples on an upward accelerating line (the
solid line). The apples were initially 
at the position of the unfilled apple, one at rest and the other
moving horizontally to the right.
Both apples have to move up
the same amount for a given coordinate time. But the one that moves
horizontally has less proper time to do it. It must therefore experience a
greater acceleration.} 
\label{fig3}
\end{center} 
\end{figure}

To further clarify the main idea, we can also consider a similar scenario
where there are two apples on a horizontal straight line which accelerates
upward relative to an inertial system, as depicted in \fig{fig3}.

It follows from the equivalence principle (see
\rapp{extraapp}) that a flat platform on Earth (neglecting the Earth's
rotation) behaves like a flat platform with proper upward acceleration $g$
in special relativity. Hence for a sufficiently flat platform, the force
required to hold an apple at rest inside a moving train on Earth would
increase as the velocity of the train increases. 

\section{The static black hole}\label{sec3}
\begin{figure}[h]
\begin{center}
\psfrag{g}{$g$}
\epsfig{figure=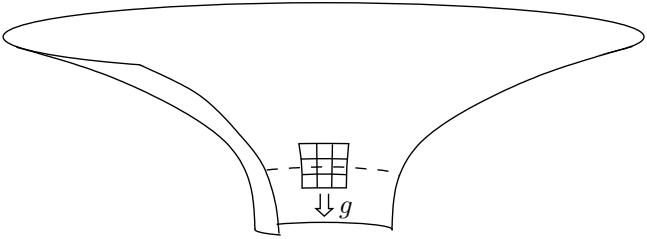,width=8.5cm}
\caption{A freely falling frame (the grid) accelerating relative
to the spatial geometry of a black hole.\supcite{parabola} We consider circular
motion along the dashed line. The bottom edge of the depicted surface 
corresponds to the horizon. At this edge the embedding
approaches a cylinder and the circle 
at the horizon is thus straight in the sense that it
does not curve relative to the surface.} 
\label{fig4}
\end{center} 
\end{figure}

Let us apply the reasoning of Sec.~\ref{train} to circular motion
around a static black hole. A schematic 
of the exterior spatial
geometry of an equatorial plane through a black hole is depicted in
\fig{fig4} (also see 
\rapp{extraapp}).

\begin{figure}[h]
\begin{center}
\epsfig{figure=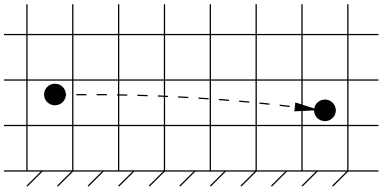,width=5cm}
\caption{A zoom-in on the circular motion observed from a static
system (with proper upward acceleration). The 
trajectory curves slightly downward, which will decrease the upward
acceleration of the object relative to a freely
falling system. In the limit that the acceleration of the freely falling
frames is infinite, we can disregard the small curvature.} 
\label{fig5}
\end{center} 
\end{figure}

A local static reference system outside of the black hole
will behave as an accelerating reference frame (train platform) in
special relativity (again see \rapp{extraapp}). Locally, the scenario is
thus identical to that in Sec.~II, except that the path along which the
object in question moves (a circle in the latter case) is not straight in
general (although circles can in fact be straight, see \rapp{extraapp}).
Instead, the circular motion corresponds locally to letting
the object in question follow a slightly curved path relative to the
accelerating platform (see \fig{fig5}).

It is a well known property of Schwarzschild black holes that the 
proper acceleration of the local static reference frame
goes to infinity as the radius approaches the radius of the event
horizon.
In other words the acceleration of a freely falling  inertial frame
(where special relativity holds, see \rapp{extraapp})
which falls relative to the static reference frame, goes to infinity as the radius approaches the radius
of the horizon. 
Furthermore we know that there is a maximum velocity
$v=1$ for material objects. Thus the perpendicular acceleration relative to
the properly accelerated reference frame due to the curvature of the path
remains finite (it is given by $v^2/R$) for non-zero $R$; $R$ is non-zero for
the circular motion in question. It follows that the acceleration
$a\su{rel}$, relative to a freely falling frame, of an object in circular
motion is dominated by the acceleration of the freely falling frame in the
limit where the radius of the circle approaches the radius of the event
horizon. Thus in this limit we can neglect the curvature of the path and
from the reasoning in \rsec{train} we conclude that the force required to
keep an object in a circular orbit (given by $F=m \gamma^2 a\su{rel}$)
increases as the orbital velocity increases.

In brief, if an object moves it has less time (due to time
dilation) to accelerate the necessary distance upward needed to remain
at a fixed height (that is, fixed radius). Thus we can understand that
that close to the horizon a greater outward force is needed to keep an object in
orbit the faster the object moves.

\section{A more quantitative analysis}\label{quant}
To understand where the transition from a more Newtonian-like 
behavior occurs, we need a more detailed analysis.
Because the reference frame connected to the static observers around the
black hole locally behaves like an accelerating reference frame in
special relativity, we first consider this special relativistic
case. 

\begin{figure}[h]
\begin{center}
\psfrag{gv}{$g_\perp$}
\psfrag{g}{$g$}
\psfrag{v}{${\bf v}$}
\epsfig{figure=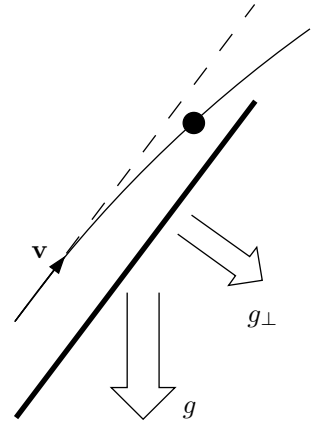,width=4cm}
\caption{A particle moving along a trajectory of curvature $R$
relative to the accelerating reference frame. The thick line is freely
falling and is initially ($t=0$) aligned with the dashed line. 
Concerning forces perpendicular to the direction of motion, only the
perpendicular part of the acceleration ${\bf g}$ is relevant.} 
\label{fig6}
\end{center} 
\end{figure}

Let ${\bf v}$ be the velocity of a particle relative to the accelerated
reference frame and let ${\bf g}$ be the acceleration of an inertial
frame, momentarily at
rest relative to the reference frame, which falls relative to the 
the reference frame. 
Also assume that the direction of
curvature $\nhat$ (discussed in \rapp{extraapp}) of the particle trajectory
relative to the reference frame 
lies in the same plane as that spanned by
${\bf v}$ and ${\bf g}$. In this way we consider an effectively
two-dimensional scenario as depicted in \fig{fig6}.

The perpendicular acceleration of a particle moving on a curve of
radius $R$, as observed from the accelerated reference frame,
is given by $v^2/R$. In other words, the proper
spatial distance $\delta s$ from the particle to a straight line fixed to
the accelerated reference frame and aligned with the particle
initial ($t=0$) direction of motion is given by
$\delta s=(\delta t)^2
v^2/(2R)$ to lowest nonzero order in $\delta t$; the latter is the local
time as measured in the accelerated reference frame. Because the inertial
(freely falling) system is initially at rest with respect to the
accelerating reference
frame, the time as measured by a grid of ideal
clocks in the freely falling system is identical to time (to first order in
$\delta t$) relative to the accelerating reference frame. The same goes for distances
(length contraction does not kick in until the two frames have an
appreciable relative velocity).

Consider a straight line fixed to the freely falling
system that at $t=0$ coincides with the previously mentioned line fixed
to the reference frame. The separation between the two lines is (to
lowest nonzero order in $\delta t$) given
by $g_\perp (\delta t)^2/2$. Here $g_\perp$ is the part of the
acceleration of the freely falling system 
that is perpendicular to the initial direction of
motion, as observed from the accelerating reference frame.
It follows that observed from the freely falling system, where special
relativity holds, the particle will have an acceleration perpendicular to
the direction of motion given by $g_{\perp}-v^2/R$. 
In analogy to our earlier reasoning, the 
perpendicular acceleration as observed in a system comoving with the
particle is greater by a factor of
$\gamma^2$, and the perpendicular force $F_\perp$ (as experienced in
the particle's own system) 
is thus given by
\begin{equation}
\label{net}
F_\perp=m \gamma^2 (g_{\perp}-v^2/R).
\end{equation}
To clarify any sign ambiguities we rewrite \eq{net} in terms of vectors:
\begin{equation}
\label{net2}
\frac{{\bf F}_\perp}{m}=-\gamma^2 {\bf g}_{\perp} + \gamma^2 v^2
\frac{\nhat}{R}.
\end{equation}
Equation~\eqref{net2} relates the perpendicular part of the force (as
observed in the particle's own reference system) to the
spatial curvature of the particle trajectory relative to a reference frame
with proper acceleration $-{\bf g}$. Although Eq.~\eqref{net2} was 
derived for an effectively two-dimensional scenario, it holds also in three
dimensions as we will see in
\rsec{3d}.
From the equivalence principle it follows that 
Eq.~\eqref{net2} applies also to motion around a black hole. For this
case the curvature vector is defined relative to the spatial
geometry connected to the static observers, as explained in \rapp{extraapp}.

\section{Following the geodesic photon}\label{gefo}
Inspired by the reasoning of Abramowicz et al.,\supcite{marek} we now consider
motion along the spatial trajectory of a geodesic photon (a photon whose
motion is determined by gravity alone, see \rapp{extraapp}). 
For a geodesic
particle we have $ F_\perp=0$, and thus according to \eq{net},
$g_{\perp}=v^2/R$. For a geodesic photon whose path curvature we denote by
$R_\script{phot}$, we have thus $g_{\perp}=1/R_\script{phot}$ (because $v=1$
for photons). 
For a particle following the path
of a such a photon (so $1/R=g_\perp$) we have according to \eq{net},
$F_\perp=m
\gamma^2 (g_{\perp}-v^2 g_{\perp})$, which simplifies to $F_\perp=m
g_{\perp}$. Thus, the perpendicular force required to make a particle follow
the trajectory of a geodesic photon is independent of the velocity
of the particle. 

To make this fact more transparent, we consider the curvature vector of
a geodesic photon given by \eq{net2} (set ${\bf F}_\perp=0$ and $v=1$)
\begin{equation}
\label{tass}
\frac{{\bf \hat{n}}\su{phot}}{R\su{phot}} = {\bf g}_{\perp}.
\end{equation}
We introduce $\nhat \su{rel}/R\su{rel}$ 
as the curvature vector relative to the trajectory of a
geodesic photon: 
\begin{equation}
\label{tiss}
\frac{{\bf \hat{n}}\su{rel}}{R\su{rel}}= \frac{ {\bf \hat{n}}}{R}-\frac{{\bf
\hat{n}}\su{phot}}{R\su{phot}}.
\end{equation}
This definition of 
$\nhat\su{rel}/R\su{rel}$ 
gives how quickly a particle trajectory deviates from a geodesic photon
trajectory in analogy to how
${\bf \hat{n}}/R$ gives how quickly the particle trajectory 
deviates from a straight line. We substitute Eqs.~\eqref{tass} and
\eqref{tiss} in
\eq{net} and find
\begin{equation}
\label{orel}
\frac{{\bf F}_\perp}{m}=-{\bf g}_\perp+\gamma^2 v^2 \frac{{\bf
\hat{n}}_\script{rel}}{R_\script{rel}}.
\end{equation}
Equation~\eqref{orel} also holds 
for more general three-dimensional scenarios as will be shown
in \rsec{3d}.

Within the photon radius, a photon trajectory departs inward relative
to a locally tangent circle.
Thus the relative curvature direction
${\bf \hat{n}}_\script{rel}$ of a circle within the photon radius is
directed outward. From \eq{orel} it then follows that the faster an object 
orbits the black hole, the greater the outward force must be. However,
outside of the photon radius, a photon trajectory departs outward from a
locally tangent circle. Thus
the relative
curvature of the circle is directed inward. It follows from \eq{orel} that
outside of the photon radius, the outward force required to keep the
rocket in orbit will decrease as the velocity increases. Thus we see that
the effective centrifugal force reversal for circular motion occurs exactly
at the photon radius. 

\begin{figure}[h]
\begin{center}
\epsfig{figure=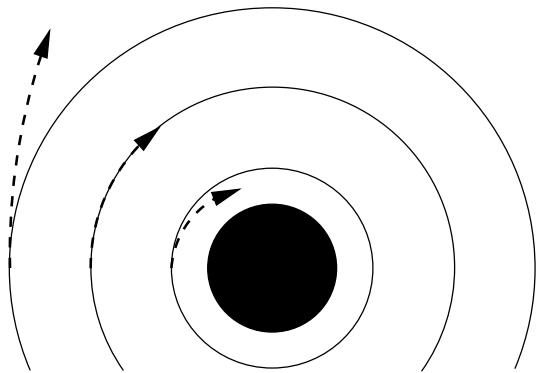,width=7cm}
\caption{Trajectories of geodesic photons (dashed curves) relative
to circles around a black hole. Inside the photon radius a
circle curves outward relative to a locally tangent photon trajectory.} 
\label{fig7}
\end{center} 
\end{figure}

\section{The difference between the\\ given and the received force}\label{givforce}

Before considering a more general
analysis, we will distinguish between two types of forces. 
The perpendicular force that we have discussed is the force as
observed in a system comoving with the object in question.
Consider now a situation where the observers connected to the
reference frame in question (like the accelerating platform we have considered previously)
provide the force that keeps the object on its path.
How is this force, which we will refer to as the {\it given} force,
related to the previously considered force, which we will refer to as the {\it received}
force, that is, the force as observed in a system comoving with the object? 
For instance, we might be interested in the magnitude of the vertical force by the rail
that is needed to support a train moving with
a relativistic speed along the track. 
Unlike in Newtonian theory, this given force will be
different from the force as observed in a system comoving with the train.

\begin{figure}[h]
\begin{center}
\psfrag{v}{$v$}
\psfrag{(a)}{(a)}
\psfrag{(b)}{(b)}
\epsfig{figure=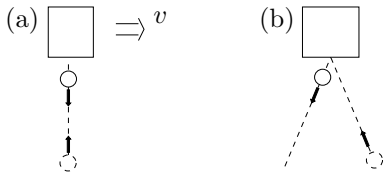,width=5cm}
\caption{A simple model where small
particles bounce elastically on the object. (a) The scenario as observed from a system where
the impulse giving particle has no horizontal velocity. (b) The corresponding scenario as observed from a system
comoving with the object.} 
\label{fig8}
\end{center} 
\end{figure}

For forces perpendicular to the direction of motion,
the relation between the given and the received force can be understood by
considering a simple model of force exertion (a more formal
derivation is given in \rcite{rickinert}).
Assume that the force on the object is
exerted by little particles bouncing elastically on the object.
Every bounce gives the object a certain impulse (see \fig{fig8}). 

If we give an object moving relative to an inertial system $S$ an impulse perpendicular to the direction of
motion, the object will in its own reference system receive the same impulse because a Lorentz transformation does not
affect the perpendicular part of the momentum change. On the other hand,
the proper time of the object runs slower by a factor of $\gamma$ compared to
local time in $S$. 
Hence the bouncing particles will bounce more frequently by a factor of
$\gamma$ as observed from the reference frame of the object. 
Because force equals transferred impulse per unit time, it follows
that the received force, perpendicular to the direction of motion, is
greater than the corresponding given force, by a factor of $\gamma$. 
We let $F_\perp$ denote the perpendicular received force and
$F_{\script{c}\perp}$ the perpendicular given force 
(to conform with the notation of \rcite{rickinert}) and write
\begin{equation}
F_{\script{c}\perp}=F_{\perp}/\gamma.
\end{equation}
Hence the given force is smaller than the received force by a factor of
$\gamma$. Because the received force 
required to keep an object (like an entire train) moving along a straight
horizontal line relative to a vertically accelerating reference frame is proportional to $\gamma^2$ (as discussed in
\rsec{train}), it follows that the force required by the rail to
support the train scales with a factor of $\gamma$.

In \rsec{gefo} we showed that the perpendicular received force is independent of
the velocity for an object that follows the trajectory of a geodesic photon.
Now we ask if there is a corresponding path for which the
perpendicular given force is velocity independent.
The analogue of \eq{net2} for the given force is
\begin{equation}\label{ghr}
\frac{{\bf F}_{\script{c}\perp}}{m}=-\gamma {\bf g}_{\perp} + \gamma v^2
\frac{\bf \hat{n}}{R}.
\end{equation}
We now require ${\bf F}_{\script{c}\perp}$ to be the same for an object
moving with speed $v$ as that of an object at rest. 
For $v=0$, \eq{ghr} gives ${\bf F}_{\script{c}\perp}/m=-{\bf
g}_{\perp}$. We substitute this result into \eq{ghr} and find
\begin{equation}
\frac{{\bf \hat{n}}}{R} =\frac{{\bf g}_{\perp}}{v^2}
\Big(1-\frac{1}{\gamma}\Big)
={\bf g}_{\perp} \frac{\gamma}{\gamma+1}.
\end{equation}
This curvature depends on the velocity.\supcite{divv} 
Thus considering the given force (the force as observed from the accelerating
reference frame), 
there is no path
for which the perpendicular force is independent of the velocity.

\section{The relativistic submarine}\label{exsec}
As an application of our discussion we consider a submarine submerged in water with a density such that
it remains at rest.\supcite{submarine}  
If we take relativistic effects into
account, 
but disregard the more subtle aspects of fluid dynamics such as viscosity and turbulence, 
will the submarine sink or float when it is given a horizontal velocity?

\subsection{A flat ocean in special relativity}\label{newsub}
We first consider a special relativistic scenario where the
flat bottom of the ocean has a constant proper upward acceleration. 
In \rcite{submarine} accelerated (Rindler) coordinates are used to
find out whether the submarine sinks or floats, 
after several pages of calculation.

\begin{figure}[h]
\begin{center}
\psfrag{v}{$v$}
\psfrag{(a)}{(a)}
\psfrag{(b)}{(b)}
\epsfig{figure=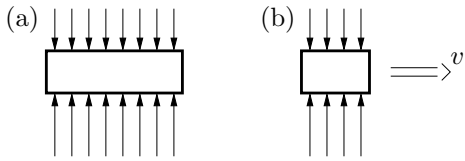,width=6cm}
\caption{(a) An idealized (rectangular) submarine submerged in water
at rest relative to the water. (b) As the
submarine moves, it will be length contracted and thus the
given force from the water will decrease by a factor of $\gamma$.} 
\label{fig9}
\end{center} 
\end{figure}

By using our earlier reasoning, we can readily find the
answer without any calculations. The received force needed to keep the
submarine at the same depth increases by a factor of $\gamma^2$ as
demonstrated in \rsec{train}. The given force needed to keep it at a
constant depth thus increases by a factor of $\gamma$ because it is smaller
than the received force by a factor of
$\gamma$ as explained in \rsec{givforce}. However, the submarine is length
contracted, so the actual given force from the water pressure (or rather the
differences of water pressure at the top and bottom of the submarine) will 
decrease by a factor of $\gamma$ (see \fig{fig9}).

\begin{figure}[h]
\begin{center}
\epsfig{figure=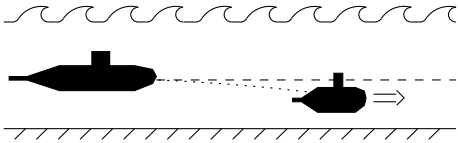,width=6cm}
\caption{A submarine submerged in a balanced state of rest in a flat
  ocean with proper upward acceleration, will sink due to
  relativistic effects if it starts moving horizontally.
} 
\label{fig10}
\end{center} 
\end{figure}

Therefore the given force decreases by a factor of $\gamma$, whereas it
should increase by a factor of $\gamma$ in order for the submarine to remain
at a fixed depth. Thus the submarine will sink (see \fig{fig10}).

\vspace{-3mm}

Now let us analyze the situation from the submarine. Due to length
contraction the actual given force is decreased by a factor of
$\gamma$, as we have argued. The received force
is $\gamma$ times the given force. Thus, the received force is independent of
the velocity.\supcite{strict}
This force
is not
sufficient to keep the submarine at the same depth. The
experienced force would have to increase by $\gamma^2$ for that.
The submarine thus sinks. 

To understand why the received force on the submarine
is independent of the velocity, we can also look at the water at the
molecular level. Assume that the
water molecules are moving along columns fastened to the ocean bottom (a
very crude model). Assume also that the particles elastically bounce back down
the same column (without interfering with the up-moving water molecules) when
they hit the hull of the submarine (and analogously for the water molecules
on top of the submarine). The impulse given by a single molecule is the same
as when the submarine was at rest. However, as observed from the moving submarine the columns of
water molecules are length contracted by a factor of $\gamma$. 
There are thus more columns under the submarine (and above) in the submarine frame,
when the submarine moves than when it is at rest. On the other hand, due to
time dilation, how often a molecule from a single column hits the hull is
decreased by a factor of $\gamma$ (consider a clock fixed to the column
just where the column intersects the submarine hull). The effects thus
cancel. It follows that the received force on the
submarine is independent of the velocity.

\begin{figure}[h]
\begin{center}
\psfrag{v}{$v$}
\psfrag{(a)}{(a)}
\psfrag{(b)}{(b)}
\epsfig{figure=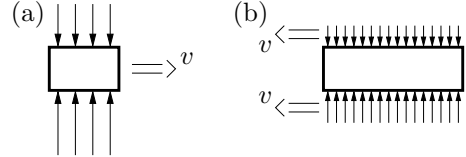,width=6cm}
\caption{(a) Observed from the water system the submarine is length
contracted by a factor of $\gamma$. (b) Observed from the submarine the
water columns are length contracted and thus denser by a factor of $\gamma$.} 
\label{fig11}
\end{center} 
\end{figure}

Although every single column of water yields a received force that is
smaller than the force given by that column (by a factor of $\gamma$), there are $\gamma^2$
times more columns contributing to the net force as observed from the
submarine frame, 
than observed from the rest frame of the water 
(see \fig{fig11}). Thus consistent with the reasoning of
\rsec{givforce}, the net received force is greater than the given force by a
factor of $\gamma$.

\vspace{-8mm}

\subsection{A real spherical ocean}
\begin{figure}[b]
\begin{center}
\epsfig{figure=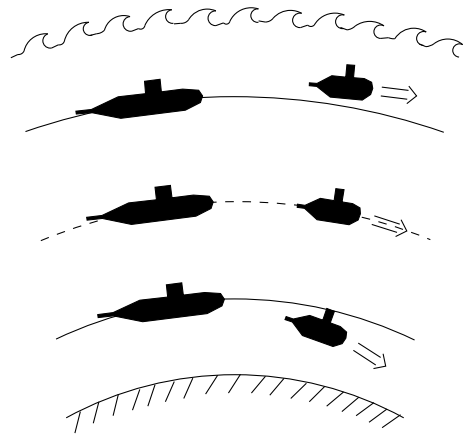,width=6cm}
\caption{Submarines moving at different depths in
the ocean of an imaginary very dense planet. 
The dashed line is the
photon radius.
The submarines
outside the photon radius will float upward if they are
given an azimuthal velocity; the opposite holds within the photon
radius.} 
\vspace{-0.3cm}
\label{fig12}
\end{center} 
\end{figure}
It is easy to generalize the above discussion to apply to a submarine
in the ocean of a spherical planet. Locally the scenario is
almost identical to the one we have discussed, assuming that the submarine is
small compared to the size of the planet. The
question of whether the submarine floats or sinks amounts to whether
the submarine, when given an azimuthal velocity, departs outward or
inward from a circle locally tangent to the direction of motion.
The answer follows from our previous
discussion.
As argued in \rsec{newsub},
the force as experienced
in a system comoving with the submarine, 
is independent of the velocity for this case. 
It then follows from \eq{orel} that the submarine will have zero curvature
relative to a geodesic photon and will thus follow the path of a
geodesic photon.
So, outside the photon radius (the radius where photons would
move on circular orbits if there were no refraction effects from the water) 
the submarine will float upward, at the photon
radius it will remain at the same depth, and inside the photon radius
it will sink. The scenario is illustrated in \fig{fig12}.

Consider the Earth, which is not sufficiently dense to have a photon
radius. If we take into account the Earth's curvature, it follows that 
when given a horizontal velocity, the submarine will not sink after all but
rather float upward.

\section{The weight of a box with\\ moving particles}
As another application of our discussion, we consider the
weight of an object whose internal components move. 
In general relativity, if we for instance heat an object, it will
become heavier. In other words, a greater upward force is required to keep the object at rest
(on Earth) when the object is warm (molecules moving faster) than when it
is cold.
Although not directly related to the main topic of this article
(inertial forces), we can give a simple explanation. 

\begin{figure}[h]
\begin{center}
\psfrag{(a)}{(a)}
\psfrag{(b)}{(b)}
\epsfig{figure=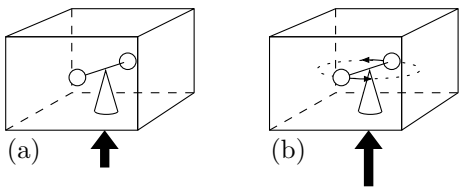,width=6cm}
\caption{A black box (transparent for 
clarity) containing a pair of balls that (a) are at
rest and (b) are moving. 
The force needed to hold
the box at a fixed height on Earth is greater when the balls are moving than
when they are at rest. The force is proportional to the
total relativistic energy
of the box.} 
\label{fig13}
\end{center} 
\end{figure}

Consider a black box containing two balls
connected by a rod of negligible mass which is suspended in such a way that the balls can rotate in
a horizontal plane. If they rotate, the upward force needed to
keep a single ball in the horizontal plane as observed from the balls' reference
system is $m g \gamma^2$, where $m$ is the rest mass of the ball. The given
force is smaller by a factor of $\gamma$ and is hence given by $m g \gamma$. So
the weight of the box is greater when the internal particles move than when
they are at rest (see \fig{fig13}).

For vertical or arbitrary motion, this type of reasoning is not as powerful,
and we can instead make a more formal proof using four-vectors
and conservation of four-momentum. 

\section{Generalizing to three dimensions}\label{3d}
\begin{figure}[h]
\begin{center}
\psfrag{x}{$x$}
\psfrag{y}{$y$}
\psfrag{z}{$z$}
\psfrag{dx1}{$\delta {\bf x}_1$}
\psfrag{dx2}{$\delta {\bf x}_2$}
\psfrag{dx3}{$\delta {\bf x}_3$}
\psfrag{g}{${\bf g}$}
\psfrag{Freely falling line}{Freely falling line}
\epsfig{figure=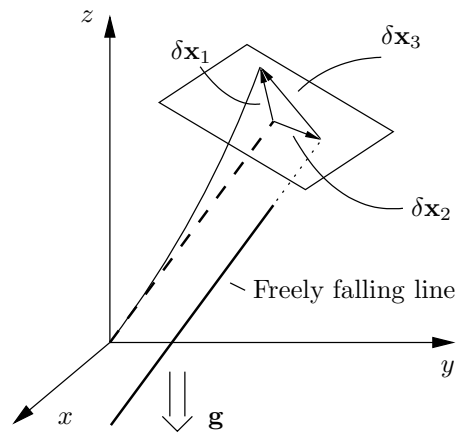,width=6cm}
\caption{Deviations from a
straight line relative to the (properly) accelerated reference system.
The $z$-direction is chosen to be antiparallel to the local 
${\bf g}$. The plane in which we study the
deviations is perpendicular to the momentary direction of motion (the dashed
line) and the three vectors lie in this plane. The solid curve
is the particle trajectory as observed in the (properly) accelerated reference 
system.
The thick line is a freely falling line that was aligned with the
dashed line (and at rest relative to the reference frame) at the time when the
particle was at the origin.
} 
\label{fig14}
\end{center} 
\end{figure}

Consider a reference frame with a proper
(upward) acceleration. 
Given the curvature and curvature direction of
the path taken by a test particle relative to the reference frame, we
want to express the
perpendicular acceleration of the test particle relative to an
inertial system $S$ in which the reference frame is momentarily
($t=0$) at rest.
In \fig{fig14} we illustrate how the trajectory will
deviate from a straight line (directed along the particle
initial direction of motion) which is fixed to $S$, and thus falls
relative to the accelerated reference frame. 
From this deviation we can find the 
perpendicular acceleration relative to $S$, analogous
to the two-dimensional discussion in \rsec{quant}.

If we let $\delta t$ denote a small time step and use the definitions
introduced in \fig{fig14}, we have to lowest nonzero order in $\delta
t$:
\begin{subequations} \label{e3}
\begin{align}
\delta {\bf x}_1&=\frac{{\bf \hat{n}}}{R} \frac{v^2 \delta t^2}{2} \\
\delta {\bf x}_2&={\bf g}_\perp \frac{\delta t^2}{2}   \\
\delta {\bf x}_3&=\delta {\bf x}_1 - \delta {\bf x}_2,
\end{align}
\end{subequations}
where $R$ and $\nhat$ are the curvature and curvature direction of the
spatial trajectory relative to the accelerated reference frame. 
Let ${\bf a}_{\script{rel} \perp}$ 
be the acceleration of the test
particle perpendicular to the direction of motion relative to the freely
falling frame. By using $\delta {\bf x}_3={\bf a}_{\script{rel} \perp}
\delta t^2/2$ and \eq{e3}, we find
\begin{equation}
\label{pop}
{\bf a}_{\script{rel} \perp}=-{\bf g}_\perp + v^2 \frac{{\bf \hat{n}}}{R}.
\end{equation}
We denote the received perpendicular force by 
${\bf F}_\perp$. According to our previous reasoning, we have ${\bf F}_\perp=m
\gamma^2 {\bf a}_{\script{rel} \perp}$, and thus
\begin{equation}
\label{aye}
\frac{1}{m \gamma^2} 
{\bf F}_\perp
= -{\bf g}_\perp + v^2 \frac{{\bf \hat{n}}}{R}.
\end{equation}
Equation~\eqref{aye} relates the experienced perpendicular
force and the curvature relative to the accelerating reference system. We note
that the only difference from its Newtonian analogue is the factor of $\gamma^2$
on the left-hand side.
In analogy to the two-dimensional discussion in \rsec{gefo},
we may introduce a curvature relative to that of a geodesic photon as
\begin{equation}
\label{aye2}
\frac{{\bf \hat{n}}_\script{rel}}{R_\script{rel}} = \frac{{\bf \hat{n}}}{R}
-\frac{{\bf \hat{n}}_{\script{phot}}}{R_{\script{phot}}}.
\end{equation}
If we use \eq{aye} to find $\hat{\bf n}_{\rm 
phot}/R_{\rm phot}$
(setting ${\bf F}_\perp=0$ and $v=1$) and substitute the expression for ${\bf \hat{n}}/R$
from \eq{aye2} into \eq{aye}, we obtain
\begin{equation}
\label{ayeo}
\frac{{\bf F}_\perp}{m}=-{\bf g}_\perp+\gamma^2 v^2 \frac{{\bf
\hat{n}}_\script{rel}}{R_\script{rel}}.
\end{equation}
We see that Eqs.~\eqref{net2} and \eqref{orel}, which were previously
derived only for effectively two-dimensional scenarios, are also valid for
arbitrary three-dimensional scenarios. For the case where the
observers at rest in the accelerating reference frame 
provide the pushing needed to keep the particle on track, we obtain the
given force as before by dividing the received force by a factor of
$\gamma$.

\section{Parallel accelerations}\label{adir}
Now that we know how the spatial curvature depends on the
perpendicular force, it would be useful also to know how
forces in the forward direction affect the speed $v$ of the particle 
relative to the accelerated reference frame.
We could derive this relation using
four-velocities,\supcite{fourcov} but for simplicity, we will use only standard results that
follow from the Lorentz-transformation.

\begin{figure}[h]
\begin{center}
\psfrag{v}{$v$}
\psfrag{v+dv}{$v+\delta v$}
\psfrag{du}{$\delta u$}
\psfrag{t0}{$t=0$}
\psfrag{tdt}{$t=\delta t$}
\epsfig{figure=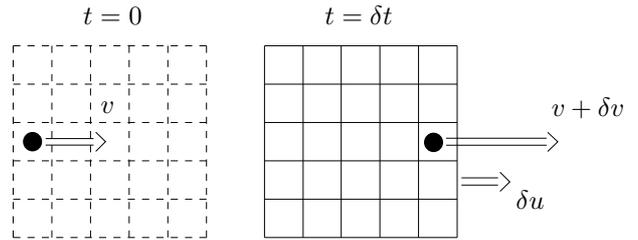,width=8cm}
\caption{An object moving relative to an accelerated reference
frame. At $t=0$ the velocity of the particle is $v$. In a time
$\delta t$ the reference frame is accelerated to
a velocity $\delta u$, and the velocity of the particle relative to
the accelerated reference frame is $v+\delta v$.} 
\label{fig15}
\end{center} 
\end{figure}

To determine $dv/dt$, where $v$ is the local velocity relative to the accelerating reference
frame, we must take into account
that the derivative implies that we are comparing the velocity at two different times,
relative to two different systems (effectively) because the reference system is accelerating.
Consider a scenario where the acceleration $a$ of the reference frame is
aligned with the direction of motion. Relative to an inertial system
$S$ in which the reference frame is at rest at $t=0$, the reference frame gains a
velocity $\delta u=a \delta t$ after a time $\delta t$. We
denote by $\delta v_s$ the velocity difference of the particle relative to
$S$ from $t=0$ to $t=\delta t$ (see \fig{fig15}).

The relation between an arbitrary object's velocity
$w$ (along the $x$-axis) as observed from $S$ and the corresponding velocity $w'$ as observed
from an inertial system $S'$ moving with velocity
$\delta u$ relative to $S$ (along the $x$-axis) follows from the
Lorentz transformation (see for example, \rcite{rindler}, p.~31)
\begin{equation}\label{neweq}
w'=\frac{w-\delta u}{1-w \delta u}.
\end{equation}
If we substitute $w = v+\delta v_s$ and $w' = v+\delta v$ in \eq{neweq}
and do a Taylor expansion to first order in $\delta u$ and $\delta v_s$, we obtain 
\begin{equation}
\label{hepp2}
\delta v=\delta v_s-(1-v^2)\delta u.
\end{equation}
We also know (see for example, \rcite{rindler}, p.\ 33), that the proper
acceleration $\alpha$ of the object, that is, the acceleration as observed in a
system comoving with the object, is related to the acceleration
$d v_s/dt$ relative to $S$ by
\begin{equation}
\label{mini}
\alpha=\gamma^3 \frac{d v_s}{dt}.
\end{equation}
If we denote the received forward thrust by $F_\parallel$, we have
$F_\parallel=m \alpha$. We use this relation in Eqs.~\eqref{hepp2} and
\eqref{mini}, take the limit where $\delta t$ is infinitesimal, together with $a=du/dt$ and find
\begin{equation}
\frac{dv}{dt}=\frac{1}{m \gamma^3} F_\parallel - \frac{a}{\gamma^2}.
\end{equation}

Consider now a more general case where the acceleration of the
reference frame need not be aligned with the direction of
motion. It is easy to realize
(or at least guess) that the
acceleration of the reference frame perpendicular to the direction of
motion will not affect the local speed derivative.\supcite{longnote} We let
${\bf g}=-{\bf a}$, where ${\bf a}$ is the acceleration of the reference
frame relative to an inertial system in which the reference frame is
momentarily at rest, and write
\begin{equation}
\label{painx}
\frac{dv}{dt}=\frac{1}{m \gamma^3} F_\parallel +
\frac{g_\parallel}{\gamma^2}.
\end{equation}
Here $g_\parallel$ is minus the part of the reference frame
acceleration that is parallel to the
particle's direction of motion. Thus we now have a general expression for the speed change relative to the
accelerating reference system. Note that $t$ is the local time
relative to the reference frame (so $dt=\gamma d\tau$).

\subsection{Combining the force equations}
From the form of Eqs.~\eqref{painx} and \eqref{aye}, we see that we can
combine them into a single vector relation. 
Let $\hat{\bf t}$ be a normalized vector in the forward direction of
motion (to conform with the notation of \rcite{rickinert}). 
By multiplying \eq{painx} by $\gamma^2 \hat{\bf t}$ and adding the resulting equation to
\eq{aye}, we can form a single term ${\bf g}$ (by adding the ${\bf g}_\perp$
and $g_\parallel {\bf \hat{t}}$ terms) and obtain
\begin{equation}
\label{pain}
\frac{1}{m \gamma^2} (\gamma F_\parallel {\bf \hat{t}} + F_\perp
{\bf \hat{m}})=-{\bf g} + \gamma^2 \frac{dv}{dt} {\bf \hat{t}} +
\frac{v^2}{R} {\bf \hat{n}}.
\end{equation}
Here ${\bf \hat{m}}$ is a unit vector perpendicular to ${\bf \hat{t}}$.
We thus have an expression for the spatial curvature and the speed
derivative in terms of the received forces. Note that $\bf{g}$ may be
interpreted as an inertial (fictitious) force; we will discuss this
interpretation in \rsec{rotating}.

We have previously considered a rocket in circular orbit with constant
speed around a black hole. Now we consider a rocket in radial motion
with constant speed outward from a black hole. From the parallel part of
\eq{pain} we find
\begin{equation}
F_\parallel=m g \gamma,
\end{equation}
where $g$ is the magnitude of the acceleration of the local freely falling
frames ($g$ is a function of the radius that can readily be found from the
spacetime metric). Here there are no reversal issues. However, we can see
that (unlike in Newtonian theory), a greater thrust is needed to keep
a constant speed the faster the rocket moves.

\subsection{The given parallel force}
If we would like an expression of the type \eq{pain} for the
parallel given force, we need to know how the given force along the direction
of motion is related to the received force along the direction of
motion. We can make an argument similar to the one we made in
\rsec{givforce}. Let $S$ denote a certain rest system, and let $S'$ be
a system in a standard (non-rotated) configuration relative to $S$, which
comoves with the object in question along the $x$-axis of $S$.
Consider the force parallel to the
direction of motion to be mediated by (very light)
particles bouncing elastically on the object. 
For simplicity let us assume that in a system comoving with the
object, each bouncing particle is reflected in such a way that the
energy of the bouncing particle is unaffected by the bounce (so $\Delta {p'}^0=0$). If
we consider motion along the $x$-axis and use the fact that the
change of momentum four-tensor transforms according to the Lorentz
transformation, we have
\begin{equation}
\Delta p^{x}=\gamma(\Delta {p'}^x + v \underbrace{\Delta {p'}^0}_{0}).
\end{equation} 
Thus the received impulse $\Delta {p'}^x$ is smaller than the given
impulse $\Delta p^x$ by a
factor of $\gamma$. On the other hand, due to time dilation the
frequency at which these impulses are received (assuming several
bouncing particles) is greater in the comoving system $S'$ than in $S$
by a factor of $\gamma$. These two
factors of $\gamma$ cancel each other, and we conclude that the given and
the received force in the direction of motion are the same. One can
easily give a formal proof of this fact (see for example,
\rcite{rickinert}). 

We can now express \eq{pain} in terms of the given
forces. We let $F_{\script{c} \parallel}$ denote the given force in the
direction of motion and write
\begin{equation}
\label{pain2}
\frac{1}{m \gamma^2} (\gamma F_{\script{c} \parallel} {\bf \hat{t}} +
\gamma F_{\script{c}\perp}
{\bf \hat{m}})=-{\bf g} + \gamma^2 \frac{dv}{dt} {\bf \hat{t}} +
\frac{v^2}{R} {\bf \hat{n}}.
\end{equation}
We define ${\bf F}_c=F_{\script{c} \parallel} {\bf \hat{t}} +
F_{\script{c}\perp} {\bf \hat{m}}$ and write Eq.~\eqref{pain2} 
as
\begin{equation}
\label{pain3}
\frac{{\bf F}_c}{m \gamma} =-{\bf g} + \gamma^2 \frac{dv}{dt} {\bf \hat{t}} +
\frac{v^2}{R} {\bf \hat{n}}.
\end{equation}
If we compare \eqref{pain3} with \eq{pain}, we see that the formalism is a bit cleaner if we
consider the given force rather than the received force.

\section{Rotating reference frame}\label{rotating}
Suppose that we would also like to consider stationary
spacetimes, such as the spacetime of a rotating (Kerr) black hole. For this case we have a
spatial geometry defined by the stationary (Killing) observers. In this case through frame
dragging, the local reference frame connected to the stationary
observers is not only accelerating, but also
rotating. 

\begin{figure}[h]
\begin{center}
\psfrag{w}{$\omega$}
\psfrag{dx}{$\delta {\bf x}$}
\epsfig{figure=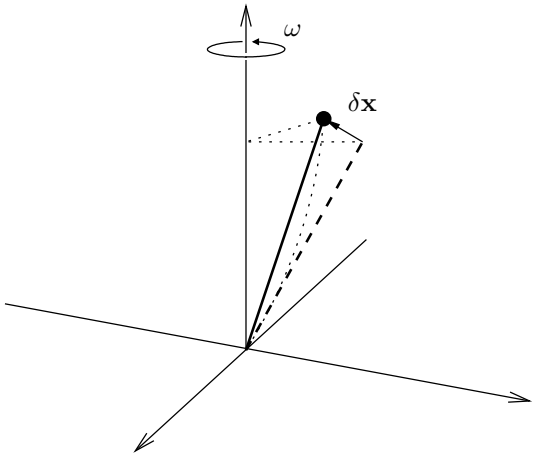,width=7cm}
\caption{A particle (the black dot) moving along a rotating straight line
(depicted at two successive time steps -- the dashed and the solid
line), as observed relative to an inertial system $S$.
Relative to $S$ the particle trajectory (the dotted line) curves.} 
\label{fig16}
\end{center} 
\end{figure}

Consider (in special relativity) a reference frame that rotates
around its origin relative to an
inertial system $S$. For simplicity, we consider motion along a
straight line that passes the origin and is fixed to the rotating
frame. The particle is assumed to be at the origin at $t=0$. This scenario is depicted in \fig{fig16}.

Let $\delta {\bf x}$ denote the perpendicular separation from the particle to
a line that is fixed in the inertial system $S$ and that at $t=0$ was aligned with the rotating line. 
We let $\delta {\bf u}$ denote the velocity of the line fixed to the
rotating system at the position of the particle after a time $\delta t$. To
lowest order in $\delta {\bf u}$ we have 
\begin{equation}
\delta {\bf x}=\delta {\bf u}\,\delta t. \label{bull}
\end{equation}
The position of the particle after a time $\delta t$ is
${\bf v} \delta t$ (to lowest order in $\delta t$), where ${\bf
v}=v {\bf \hat{t}}$. We thus have $\delta {\bf u}=\fat{\omega} \times ({\bf
v} \delta t)$. We use this relation in \eq{bull} and obtain 
$\delta {\bf x}=\fat{\omega} \times {\bf v} \delta t^2$. In the limit where the
time step is infinitesimal, the perpendicular acceleration coming from the rotation
is
\begin{equation}
{\bf a}_{\script{rel} \perp}= 2 \fat{\omega} \times {\bf v}.
\end{equation}
For the low reference frame velocities that occur during the
short time $\delta t$, the effects of length contraction and time dilation
will not enter the expressions for the perpendicular deviations (to
lowest nonzero order). 
Therefore we can add the effect of rotation to the effects of curvature and acceleration.
The generalization of \eq{pop} is thus
\begin{equation}
{\bf a}_{\script{rel} \perp}=-{\bf g}_\perp + 2 \fat{\omega} \times {\bf v}+ v^2 \frac{{\bf
\hat{n}}}{R}.
\end{equation}
Here ${\bf a}_{\script{rel} \perp}$ is the perpendicular acceleration of the test
particle relative to the inertial system $S$.

Because the changes in the reference frame velocity (as observed from the inertial
system in question) are perpendicular to the direction of motion,
the derivative of the speed will not be affected by the rotation. 
Thus we can write the generalization of \eq{pain} as 
\begin{equation}
\label{paingen}
\frac{1}{m \gamma^2} (\gamma F_\parallel {\bf \hat{t}} + F_\perp
{\bf \hat{m}})=-{\bf g} + 2 \fat{\omega} \times {\bf v}+ \gamma^2
\frac{dv}{dt} {\bf \hat{t}} +
\frac{v^2}{R} {\bf \hat{n}}.
\end{equation}
Equation~\eqref{paingen} relates the real received forces to
both the curvature and the speed change per unit time relative to the
accelerating and rotating reference
frame. 
Note that although ${\bf g}$ is minus the acceleration of the reference
frame, $\fat{\omega}$ is the rotation vector of the reference frame.

As an application we consider a person walking on a
straight line through the center of a rotating flat merry-go-round
(in special relativity). 
The perpendicular force experienced as he/she passes the center (where ${\bf g}$ is zero) is 
given by \eq{paingen} as
\begin{equation}
\label{30}
F_\perp=2\omega_0 v \gamma^2.
\end{equation}
Here $\omega_0$ is the angular frequency of the merry-go-round.
Apart from the $\gamma^2$ factor, Eq.~\eqref{30} is the same as the
corresponding equation in Newtonian mechanics. For points other than
the central point we must consider that the proper rotation $\omega$
(as measured by an observer riding the
merry-go-round at the point in question) is different from the
rotation $\omega_0$ as observed from the outside.\supcite{rotcom}

\section{Discussion}
On the left-hand side of \eq{paingen} there are real forces as
experienced in a system comoving with the object in question.
On the right-hand side the first two terms multiplied by $-m$ may be interpreted as inertial forces
\begin{subequations}
\begin{align}
\textrm{Acceleration:}~& {\ \ \ \ } m {\bf g}, \\
\textrm{Coriolis:}~& -2 m \fat{\omega} \times {\bf v}.
\end{align}
\end{subequations}
We might be tempted to denote the first term by ``gravity''
rather than ``acceleration,'' 
but if we consider 
a rotating merry-go-round as a
reference frame, this term would correspond to what is commonly
called the centrifugal force. To avoid confusion we therefore label this term
``acceleration.'' For the second term the name Coriolis is 
obvious in analogy with the standard notation for inertial
forces in non-relativistic mechanics. 

Note that what we call an inertial force is
ambiguous. For example, we could multiply the perpendicular part of
\eq{paingen} by $\gamma$. 
By defining ${\bf F}= F_\parallel {\bf \hat{t}} + F_\perp
{\bf \hat{m}}$, we could then simplify the left-hand side of \eq{paingen} 
to ${\bf F}/m\gamma$. However, because of the $\gamma$-multiplication
we would need to express the ${\bf g}$-term as a sum of a parallel and a perpendicular part
(with different factors of $\gamma$), thus creating two different
acceleration terms. There is is thus more than one way of expressing
\eq{paingen}, and identifying inertial forces, that reduce to the
Newtonian analogue by setting $\gamma=1$.

We do not regard the last two terms on the right-hand side of \eq{paingen} 
as inertial forces, but rather as descriptions of the motion (acceleration) relative to the frame of
reference. There are alternative interpretations; see
\rcite{rickinert} for further discussion.

Note that $dt$ is the local time (for the local reference frame
observers) and is related to the proper time $d \tau$ for the particle in
question by $dt=\gamma d \tau$. Equation~\eqref{paingen} is identical to the more
formally derived corresponding expression in \rcite{rickinert}. 

We have considered accelerating and rotating reference
frames, but not shearing or expanding reference frames. 
The extension is straightforward for an isotropically expanding reference frame, but for
brevity we refer to \rcite{rickinert}.

In summary, we have seen how we can derive a formalism of inertial
forces that applies to arbitrary rigid reference frames in special and
general relativity. Apart from factors of gamma, the formalism is locally
equal to its Newtonian counterpart. We have also applied the insights and
formalism of this paper to various examples, such as moving trains and submarines.

\appendix
\renewcommand{\theequation}{A\arabic{equation}}     
\setcounter{equation}{0}                            

\section{A comment on static spacetimes, index notation, and the
optical geometry}\label{opt}
For the purposes of this article it is not necessary to discuss
a formalism known as optical geometry.
However, because the latter is the
inspiration for this article and the formalisms are
very similar, a comment is in order. The index formalism 
(which distinguishes between covariant and contravariant vectors) is vital for
the comparison. 

Suppose that we have a static spacetime with the line element
\begin{equation}\label{line}
ds^2 = -e^{2\Phi}dt_\script{c}^2 +g_{ij}dx^i dx^j.
\end{equation}
We denote coordinate time by $t_\script{c}$ so as not to confuse it
with the local time of the reference frame which we denote by
$t$. Also, Latin indices are spatial indices running from 1--3. 
It is easy to show (see for example, \rcite{rickinert}, Appendix E) 
that the acceleration of the freely falling frames for a line element
of this form is given by ${\bf g}=-\nabla \Phi$. We can equivalently
write this relation as
$g^k=-g^{kj}\nabla_j \Phi$. For later convenience we define
$F_\perp^k=F_\perp m^k$, where $m^k$ is a normalized spatial vector.
If we use these results and
definitions, we can rewrite \eq{ayeo} as
\begin{equation}\label{ayeoi}
\frac{F_\perp}{m} m^k=[g^{kj}\nabla_j \Phi]_\perp +\gamma^2 v^2 \frac{{n}^k_\script{rel}}{R_\script{rel}}.
\end{equation}
Here $\perp$ means that we should select the part perpendicular to the
spatial direction of motion $t^k$.
For a line element such as \eq{line}, the optical geometry (see for example,
\rcite{optiskintro} although a different
sign convention for $\Phi$ is used)
is given by a rescaling of the standard spatial geometry
\begin{equation}
\tilde{g}_{ij}=e^{-2\Phi} g_{ij}.
\end{equation}
We thus stretch space by a factor $e^{-\Phi}$ to create a new
spatial geometry. We may consider both metrics to live on the same
(sub)manifold. 
Relative to the rescaled geometry, the curvature of a given spatial
(coordinate) trajectory is in general different from that relative to
the standard spatial geometry.
In particular, the spatial trajectories of
geodesic photons are straight with respect to the optically rescaled
space. 
It follows 
that the curvature and curvature direction with
respect to the rescaled (optical) space gives how fast (with respect to
the distance along the trajectory) and in what direction a
trajectory deviates from that of a geodesic photon. This curvature
and curvature direction
thus correspond to the relative
curvature and curvature direction introduced in \rsec{gefo} and \rsec{3d}, except
that the deviation and the distance along the trajectory are now rescaled.
The optical spatial curvature $\tilde{R}$,
the optical curvature direction ${\tilde{n}}^k$, and the
optically normalized direction of the perpendicular force
${\tilde{m}}^k$ for a certain
(coordinate) trajectory are related to $R_\script{rel}$,
${n}^k_\script{rel}$, and ${m}^k$ by\supcite{reason}
\begin{subequations}
\begin{align}\label{aa}
{\tilde{R}}&=e^{-\Phi} R_\script{rel}\\
{{\tilde{n}}^k}&=e^{\Phi} {{n}^k_\script{rel}}\\
{{\tilde{m}}^k}&=e^{\Phi} {{m}^k}.
\end{align}
\end{subequations}
If we use $\tilde{g}^{ij}=e^{2\Phi} g^{ij}$, we may 
rewrite \eq{ayeoi}
as (multiply the entire expression by $e^{2\Phi}$)
\begin{equation}
\label{a1}
\frac{F_\perp}{m} e^{\Phi}\tilde{m}^k=[\tilde{g}^{kj}\tilde{\nabla}_j
\Phi]_\perp +\gamma^2 v^2 
\frac{{\tilde{n}}^k}{\tilde{R}}.
\end{equation}
Note that because the covariant derivative acts on a scalar (in contrast
to a vector for example), we have
$\tilde{\nabla}_j=\nabla_j \Phi$ 
(although $\tilde{\nabla}^j \Phi=e^{2\Phi}{\nabla}^j \Phi$).
By comparing \eq{a1} with the more general (and more
formally derived) corresponding equation in \rcite{rickinert}, we have
a perfect match.\supcite{rickcom1} 
In covariant form (lower indices with $\tilde{g}_{ij}$) \eq{a1} becomes slightly
more compact:
\begin{equation}
\label{a2}
\frac{F_\perp}{m} e^{\Phi}\tilde{m}_k=[\tilde{\nabla}_k \Phi]_\perp
+\gamma^2 v^2 \frac{{\tilde{n}}_k}{\tilde{R}}.
\end{equation}
Because ${{\tilde{m}}_k}=e^{-\Phi} {m}_k$, 
the left-hand side of \eq{a2} can be expressed as $F_{\perp
k}/m$. On the other hand, the left-hand side of \eq{a1} can be
written as $F_{\perp}^k e^{2\Phi}/m$. 
Expressed in these forms, but using the boldface vector notation, the right-hand
sides of \eq{a1} and \eq{a2} are identical and the left-hand sides differ
by a factor $e^{2\Phi}$.
We hence understand the hazard of using the bold face vector
notation, at least if we use vectors that naturally ``belong'' to
two different metrics in the same expression. 
As Eqs.~\eqref{a1} and \eqref{a2} are written, only vectors
belonging to the optical geometry are used, and we could use
vector notation after all.

The parallel part of \eq{pain} in index notation (for the line element
in question and a static reference frame) takes the form
\begin{equation}\label{parallel}
\frac{1}{m \gamma} F_\parallel t^k= [g^{kj} \nabla_j \Phi]_\parallel +
\gamma^2 \frac{dv}{dt} t^k.
\end{equation}
Here we have $dt=e^\Phi dt_\script{c}$. We use the latter relation, rewrite the tensors in terms of their rescaled analogues, multiply the
entire expression by $e^{2\Phi}$, and add it to \eq{a1}. The result is
\begin{equation}
\label{parallel2}
\frac{e^\Phi}{m}\Big(\frac{F_\parallel}{\gamma} \tilde{t}^k + F_\perp
\tilde{m}^k \Big)
= \tilde{g}^{kj}\tilde{\nabla}_j \Phi + \gamma^2
\frac{dv}{dt_\script{c}} \tilde{t}^k + \gamma^2 v^2
\frac{{\tilde{n}}^k}{\tilde{R}}.
\end{equation}
Equation~\eqref{parallel2} is the inertial force formalism in terms of
the optical geometry. Again it agrees
with the corresponding equation of \rcite{rickinert}.\supcite{rickcom2} 

\section{Some basic concepts}\label{extraapp}
This appendix is included for readers with little or no background in
differential geometry or Einstein's theory of gravity.

\begin{figure}[h]
\begin{center}
\psfrag{n1}{${\bf \hat{n}}$}
\psfrag{R}{$R$}
\epsfig{figure=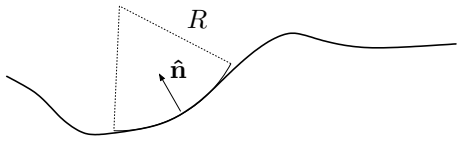,width=6cm}
\caption{A path on a plane always corresponds locally to
a circle as far as direction and curvature are concerned.} 
\label{fig17}
\end{center} 
\end{figure}

{\it Curvature}. Consider a curved path on a plane. At any point along the
path we can find a circle that is precisely tangent to the path and whose
curvature matches that of the path (see Fig.~\ref{fig17}). At any point along the curve we can thus introduce a curvature
direction ${\bf \hat{n}}$, a unit vector, and a
curvature radius $R$ as shown. The greater the curvature, the
smaller the curvature radius. 
For paths that are not in a plane
we can locally match a circle to every point along the path and define
the curvature direction and curvature radius analogously. Note that the
curvature direction ${\bf \hat{n}}$ is always perpendicular to the path.

\begin{figure}[h]
\begin{center}
\epsfig{figure=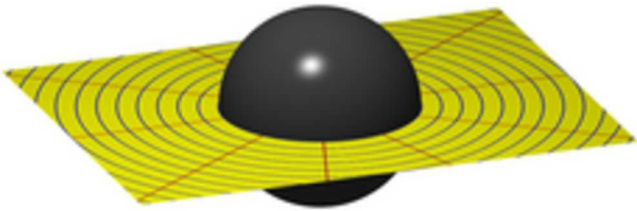,width=8.4cm}
\caption{A symmetry plane through a black hole.}
\label{fig18}
\end{center} 
\end{figure}

{\it Spatial geometry}. Consider a symmetry plane through a black hole. For
the purposes of this article we may illustrate the black hole as 
a black sphere (see \fig{fig18}). 

\begin{figure}[h]
\begin{center}
\epsfig{figure=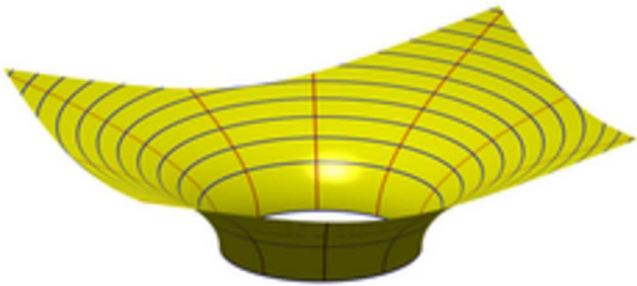,width=8.4cm}
\caption{\label{fig19}Sketch of the apparent geometry of a symmetry
plane through a black hole. The innermost circle is at the surface of the
black hole.}
\end{center} 
\end{figure}

If we could walk around on the plane and measure distances, we would
notice that the distances 
would not match those we would expect from a flat plane. Rather,
the apparent geometry would be as that depicted in Fig.~\ref{fig19}.
In particular, we would note that as one walks outward from the surface
of the black hole, the circumference would initially hardly change. Although the geometry of the curved surface corresponds to the
geometry of the symmetry plane, the symmetry plane neither curves
upward nor downward in reality. Distances on the plane
are {\it as if} the plane curves as depicted in \fig{fig19}.

{\it Straight lines as geodesics}. On a curved surface we can determine if
a line is straight or curved at a certain point by looking at the
line. We position our eye somewhere on an imagined line extending from
the point in the direction of the normal to the surface,
and look down along this imagined line at the surface.
If the line on the surface looks straight, it is straight. If the line looks
curved, it is curved. A line that everywhere, as seen from the local normal,
looks straight, is known as a {\it geodesic}. For a
spherical surface like the surface of the Earth, the equator is a geodesic.

For a line that is not straight, we can introduce a curvature direction and a
curvature radius by considering how fast and in what direction the line deviates
from a corresponding straight line on the surface,
analogous to the definition for flat surfaces.

In Einstein's theory of relativity, the motion of particles whose
motion is determined by gravity alone
corresponds to geodesics in curved spacetime.
For the purposes of this article it is sufficient to know that a
geodesic particle 
is a particle that is free to move as gravity alone
dictates. Examples are a dropped apple or a flying cannonball (assuming that we neglect air resistance). In general relativity there is no gravitational force, but there are
forces such as air resistance. These forces cause objects to deviate from the 
motion determined by gravity.

\begin{figure}[h]
\begin{center}
\psfrag{(a)}{(a)}
\psfrag{(b)}{(b)}
\epsfig{figure=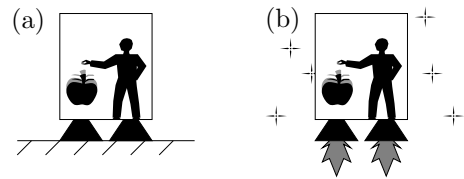,width=6cm}
\caption{
Dropping an apple inside an elevator on Earth gives the same
motion relative to the elevator as dropping it inside a (properly) accelerated 
elevator in outer space. In both cases we can introduce an inertial
(fictitious) gravitational force -- but there is (in either case) no
real gravitational force (in Einstein's theory). 
}
\label{fig20}
\end{center} 
\end{figure}

The {\it equivalence principle} can be formulated as follows:
At any point in space and time we can introduce freely falling coordinates
relative to which special relativity holds. As an example we consider an elevator whose support cables have just
snapped at the topmost level of a high building. An observer dropping a
coin inside the elevator will note that the coin will float in front of
him. If he tosses the coin, he will note that the coin moves away from him
on a straight line with constant speed just as it would if the
elevator was in outer space where there is no gravity and special
relativity holds. 
We can alternatively say that being in an elevator at rest on Earth
is equivalent to being in an
accelerated elevator in outer space (see \fig{fig20}).

\begin{figure}[h]
\begin{center}
\epsfig{figure=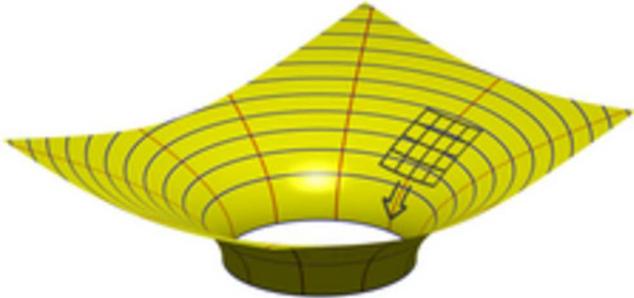,width=8.4cm}
\caption{A coordinate system 
accelerating (falling) relative to the curved spatial geometry of a black hole.}
\label{fig21}
\end{center} 
\end{figure}

It is a standard technique of Einstein's general theory of relativity
to first understand how a scenario will work relative to a freely
falling frame where everything is simple, and then
express the result with respect to the coordinates that
really interest us. These freely falling frames
are however not falling relative to a flat spatial geometry.
For the particular case of a symmetry plane of a static black hole (see
\fig{fig18}), we can 
imagine the freely falling frames (a coordinate grid in this case) to be
falling relative to the curved geometry depicted in \fig{fig19}. 
How fast the freely falling frames accelerate depends on the position (the
radius). At spatial infinity the acceleration is zero and at the
horizon it is infinite. 
The idea is illustrated in \fig{fig21}.

The depicted freely falling frame coordinate lines are
geodesics\supcite{notquite} on the curved surface. With respect to the falling
coordinate grid a free particle, that is, a particle whose motion is
determined by nothing but gravity, will move in a straight line. This
law of motion applies to all free particles, including free photons.
Because the freely falling
system is accelerating relative to the spatial geometry, the paths of free
particles will curve relative to the spatial geometry. The fact that the spatial geometry is curved does not
complicate the analysis
as far as this paper is concerned. 
The point is that locally we can always consider the geometry to be flat.
Living on a small patch of the curved surface is like living in an
accelerated reference system in special relativity. It is
only when we consider circles around the black hole that we need to
think about the spatial geometry to determine the correct curvature
of the circular path. For instance, due to the curved spatial geometry,
the innermost circle (at the surface of the black hole) is not curved at all.


\begin{thebibliography}{999}

\bibitem{hori}For the purposes of this article we
may consider the event horizon to be an (invisible) sphere. If one
ventures inside of this sphere, one cannot come back out again.

\bibitem{photradius}{There is a radius where free photons, that is, photons
whose motion is determined only by gravity, can move on circular
orbits around a black hole. The circumference of this circle is 1.5 times the circumference
of the surface of the black hole (the event horizon).}

\bibitem{marek}Marek A. Abramowicz, ``Relativity of inwards and outwards:
An example,'' Month. Not. Roy. astr. Soc. {\bf 256}, 710-718 (1992).

\bibitem{orig}Marek A. Abramowicz and Jean-Pierre Lasota, ``A note of a
paradoxical property of the Schwarzschild solution,'' Acta Phys. Pol. {\bf
B5}, 327-329 (1974).

\bibitem{carter}Marek A. Abramowicz, Brandon Carter, and Jean-Pierre Lasota, 
``Optical reference geometry for stationary and static dynamics,''
Gen. Relativ. Gravit. {\bf 20}, 1173--1183 (1988).

\bibitem{mnras1}Marek A. Abramowicz and A. R. Prasanna, ``Centrifugal force
reversal near a Schwarzschild black-hole,'' 
Mon. Not. R. Astr. Soc. {\bf 245}, 720--728 (1990).

\bibitem{mnras3}Marek A. Abramowicz, ``Centrifugal force: A few
surprises,'' Mon. Not. R. Astr. Soc. {\bf 245}, 733--746 (1990).

\bibitem{nature}Bruce Allen, ``Reversing centrifugal forces,'' Nature {\bf
347}, 615--616 (1990).

\bibitem{submarine} George E. Matsas, ``Relativistic Archimedes law for fast
moving bodies and the general-relativistic resolution of the `submarine
paradox','' Phys. Rev. D {\bf 68}, 027701-1--4 (2003).

\bibitem{rickinert}Rickard Jonsson, 
``Inertial forces and the foundations of optical geometry,'' 
Class. Quantum Grav. {\bf 23}, 1--36 (2006). 

\bibitem{proper}The proper acceleration of an object is the acceleration
measured relative to an inertial system (a freely falling system)
momentarily comoving with the object.

\bibitem{cett}Strictly speaking we are using geometrized
units in which $c=1$. In these units time has the same dimensions as
distance. If we want to express distances and times in terms of
standard units, we should replace any instance of $v$ by
$v/c$, where $c$ is the velocity of light in standard units. 

\bibitem{parabola}The embedded geometry ($t=0$ and
$\theta=\pi/2$ in Schwarzschild coordinates) corresponds to a section
of a parabola (see \rcite{gravitation}), $z=2\sqrt{R_{\script{G}}}
\sqrt{r-R_{\script{G}}}$, revolved around the vertical ($z$) axis
($R_{\script{G}}$ is the radius at the event horizon).

\bibitem{divv}In the derivation we divided by $v$, thus assuming
$v\neq0$. For $v=0$ any $\hat{\bf n}/R$ will do.

\bibitem{strict}Strictly speaking the 
argument only holds exactly as long as the velocity is purely
horizontal.

\bibitem{fourcov}Consider a $1+1$ dimensional scenario. Let $u$ be the velocity of the reference frame
relative to an inertial system in which the reference frame is
momentarily at rest, and let $\eta^\mu=(\gamma(u),\gamma(u) u)$ be the
corresponding four-velocity. Let $v_s$ be the velocity of the
test particle relative to the inertial system in question, and
let $u^\mu=(\gamma(v_s),\gamma(v_s) v_s)$ be the corresponding
four-velocity. Let $v$ be the velocity of the test particle relative
to the reference frame. We have $\gamma(v)=-\eta^\mu
u_\mu$ (using the $(-,+,+,+)$ convention). If we differentiate both sides
of this expression by
$d/dt$ and use the fact that $u=0$ and $v=v_s$ momentarily, we
find $v\frac{dv}{dt} \gamma^4=-\gamma^2 v
\frac{du}{dt}+v\frac{dv_s}{dt} \gamma^4$. This result corresponds
to \eq{hepp2} (multiply \eq{hepp2} by $\gamma^4$, divide by $\delta t$,
and take the limit where $\delta t$ is infinitesimal). 
Also we know that the proper acceleration $\alpha$ is given by
$\alpha^2=-\frac{du^\mu}{d\tau} \frac{du_\mu}{d\tau}$. 
If we differentiate $u^\mu=(\gamma(v_s),\gamma(v_s) v_s)$ with respect
to $\tau$, we find 
$\frac{du^\mu}{d\tau}=\gamma^3 \frac{dv_s}{d\tau} (v,1)$. It follows
that $\alpha=\gamma^3 \frac{dv_s}{dt}$, which is \eq{mini}.

\bibitem{rindler}Wolfgang Rindler, {\it Introduction to Special Relativity} 
(Clarendon Press, Oxford, 1982), 2nd ed.

\bibitem{longnote}It is true in the Newtonian limit that
accelerations of the reference frame perpendicular to the direction of
motion do not affect the local speed derivative. We can also
reason this 
strictly relativistically knowing a little
about time dilation and simultaneity. We could also understand it using
four-tensors knowing that $\gamma=\eta^\mu u_\mu$, where $\eta^\mu$ is the
reference frame four-velocity and $u^\mu$ is the particle four-velocity. 
Consider a particle moving in the $x$-direction and consider the
reference frame to reach a velocity $\delta v_y$ in the
$y$-direction after a time $\delta t$. To first order in $\delta t$
we have 
$\eta^\mu: (1,0,\delta v_y,0)$ and $u^\mu$: $(\gamma_0,v_x,0,0)$. 
Here $\gamma_0$ is the value of $\gamma$ at
$t=0$. We see that to first order $\gamma=\eta^\mu u_\mu=\gamma_0$, and
thus a perpendicular acceleration of the
reference grid does not affect the local speed derivative. Similarly an
infinitesimal perpendicular velocity of the {\it particle} will have no
first order effect on the speed. Any one-dimensional reasoning of how
the speed derivative is related to force parallel to the direction of
motion thus holds also when there are perpendicular effects.

\bibitem{rotcom}The difference between the proper rotation and the
rotation as observed from outside of the merry-go-round, for the
non-central points, is not only due to time dilation, but also to relativistic precession
(rotation) effects. We can also use the formalism of this paper
for these points assuming that we correctly express the proper
local reference frame rotation 
$\fat{\omega}$, that is, the rotation as experienced by an observer
at rest relative to the reference frame at the points in question. 


\bibitem{optiskintro} Marek A. Abramowicz and Jean-Pierre Lasota, ``A brief
story of a straight circle,'' Class. Quantum Grav. {\bf 14}, A23--A30 (1997).

\bibitem{reason}It is easy to make a formal proof of how the factors
should enter. For contravariant vectors one can reason it out instead.
Consider two coordinate points separated by an infinitesimal vector $dx^k$. Assume
that we stretch space by a factor
$e^{-\Phi}$. A coordinate vector that has the same norm (length) relative to
the stretched space that $dx^k$ had relative to the standard space would
have to be component-wise smaller 
by a factor $e^{-\Phi}$. 
Hence given a normalized vector relative to the standard space, we
obtain a corresponding normalized vector relative to the rescaled space by
dividing by the stretching factor $e^{-\Phi}$.

\bibitem{rickcom1}Here we consider the
perpendicular part of the spatial part of Eq. (44) of \rcite{rickinert} 
and set $\tilde{\theta}^{\alpha \beta}=0$ as is appropriate for this case.

\bibitem{rickcom2}Set $\tilde{\theta}_{\alpha \beta}=0$,
$\tilde{\eta}^\alpha
\tilde{\nabla}_\alpha \Phi=0$ and identify $\tilde{\tau}_0=t\su{c}$ in Eq.
(44) of \rcite{rickinert}.

\bibitem{notquite}Actually they are not all exact geodesics. They cannot
be in general (while at the same time being orthogonal) when the surface is
curved. But at the center of coordinates they are orthogonal, and the
curvature of the coordinate lines vanishes, which
is sufficient for the type of arguments made in this article.

\bibitem{gravitation}Charles W. Misner, Kip S. Thorne, and John A.
Wheeler, {\it Gravitation} (W. H. Freeman, New York, 1973), p. 615, Eq. (23.34b).

\end{thebibliography}
\end{document}